\documentstyle[aps,preprint,tighten,epsf,floats]{revtex}

\def\Jn#1#2#3#4{{#1} {\bf #2}, #3 (#4)}

\def\PRD{{\em Phys. Rev.} D}

\def\NP{{\em Nucl. Phys.}}

\def\PRL{{\em Phys. Rev. Lett.}}

\def\ERR{{\em erratum}}
\def\bi{\bibitem}
\newcommand{\piNN}{$\pi N\!N$}
\newcommand{\ggamma}{{\mit \Gamma}}

\newcommand{\dG}{\delta {G}}
\newcommand{\dK}{\delta {K}}
\newcommand{\GG}{G}
\newcommand{\K}{K}
\newcommand{\be}{\begin{equation}}
\newcommand{\ee}{\end{equation}}
\newcommand{\eqn}[1]{\label{#1}}
\newcommand{\eq}[1]{Eq.~(\ref{#1})}

\newcommand{\fign}[1]{\label{#1}}
\newcommand{\fig}[1]{Fig.~\ref{#1}}
\newcommand{\bpsi}{\bar{\psi}}

\begin{document}

\title{Complete set of electromagnetic corrections \\ 
to strongly interacting systems}
\author{A. N. KVINIKHIDZE,\footnote{On leave from The
Mathematical Institute of Georgian Academy of Sciences, Tbilisi, Georgia.}
B. BLANKLEIDER }

\address{Department of Physics, The Flinders University of South Australia,\\
Bedford Park, SA 5042, AUSTRALIA\\E-mail: kvinikh@daria.ph.flinders.edu.au}


\maketitle

\begin{abstract}
We show how to obtain a complete set of electromagnetic corrections to a given
nonperturbative model of strong interactions based on integral equations.  The
gauge invariance of these corrections is a consequence of their completeness.
\end{abstract}


\section{Introduction}

Electromagnetic corrections to strong interaction models play a crucial role in
a number of important problems in nuclear and particle physics. Together with
the $u$-$d$ quark mass difference, electromagnetic corrections are responsible
for isospin symmetry breaking and charge symmetry breaking (CSB) in hadronic
systems. A manifestation of this for strongly bound hadronic systems is the mass
splittings of isomultiplets as for example in the case of the neutron-proton
mass difference \cite{stephenson} and the binding energy difference of $^3$H and
$^3$He \cite{sasakawa}. Similarly for hadron-hadron scattering the
electromagnetic corrections play an important role in causing isospin symmetry
and charge symmetry breaking; indeed there has been an intensive effort at the
meson factories to measure CSB in the reactions $\pi N\rightarrow \pi N$,
$np\rightarrow np$, $np\rightarrow d\pi^0$, $\pi^\pm d\rightarrow \pi^\pm d$,
etc. \cite{miller}. Electromagnetic corrections also play an important role in
extracting strong interaction quantities from experiments that are sensitive to
the presence of the electromagnetic interaction.  A current example is the
proposed determination of the $\pi$-$\pi$ scattering lengths from measurements
of the lifetime of $\pi^+$$\pi^-$ atoms \cite{rusetsky}. Although the
instability of this atom is due to the strong interaction process
$\pi^+\pi^-\rightarrow\pi^0\pi^0$, the exact lifetime is sensitive to
electromagnetic corrections.

The most significant electromagnetic corrections are those of lowest order
($e^2$) in the electromagnetic coupling constant $e$.  Yet because of the
nonperturbative nature of strong interactions, it has not been known how to
include electromagnetic interactions in such a way that {\em all} possible
lowest order corrections are included. Without the ability of calculating the
complete set of lowest order corrections, any estimates of isospin violation,
CSB, or other relevant quantity of interest cannot be considered to be reliable.

In this paper we solve this problem by showing how to include the complete set
of lowest order electromagnetic corrections into a nonperturbative strong
interaction model based on integral equations.

\section{Electromagnetic correction to the Green function}

We work in the framework of relativistic quantum field theory where the strong
interactions of a quark or hadronic system are described by a model potential
$K$.  For clarity of presentation we treat all particles as distinguishable;
consideration of identical particle symmetry does not affect the
method for including electromagnetic interactions presented here and will be
discussed elsewhere.  Thus the strong interaction Green function $G$ is modelled
by solving the integral equation
\be
G=G_0+G_0 K G  .     \eqn{G}
\ee
\eq{G} is a symbolic equation where momentum, spin, and isospin variables have
been suppressed, and where integrations over intermediate state momenta and sums
over internal spins and isospins are also not shown. We shall use such a
shorthand notation throughout.

To obtain the complete set of lowest order electromagnetic corrections to the
strong interaction Green function $G$, we need to insert an internal photon with
propagator $D_{\mu\nu}$ into $G$ in all possible ways. This can be achieved
even though $G$ is given nonperturbatively by a simple extension of the
gauging of equations method introduced in Ref.\ \cite{g4d}. In this
method \eq{G} is used to write 
\be
G^\mu = G_0^\mu + G_0^\mu K G + G_0 K^\mu G + G_0 K G^\mu    \eqn{G^mu_eq}
\ee
where $G^\mu$ is the quantity obtained from $G$ by attaching an external photon
everywhere in the strong interaction contributions to $G$ (the quantities
$G_0^\mu$ and $K^\mu$ are similarly obtained from $G_0$ and $K$ by attaching
photons everywhere). \eq{G^mu_eq} follows from the requirement of coupling an
external photon everywhere in \eq{G}.  Similarly the requirement of coupling an
{\em internal} photon everywhere in \eq{G} allows us to write down the equation
for the complete set of lowest order electromagnetic corrections to $G$:
\be
\dG=\dG_0+\dG_0KG+G_0\dK G+G_0K\dG
+\left(G^\mu_0 K^\nu G+
G^\mu_0K G^\nu+G_0K^\mu G^\nu\right)D_{\mu\nu}  \eqn{dG}
\ee
where $\dG$ and $\delta G_0$ are the complete set of lowest order
electromagnetic corrections to $G$ and $G_0$ respectively.  Unlike $\dG$ which
has internal photons coupled everywhere, $\dK$ consists of the strong
interaction potential $K$ with all possible photon insertions {\em except} those
that have one or both photon legs attached to an external line. This should be
clear from \eq{dG} where such terms, missing from $\dK$, are included in other
terms of the equation.  Note that the last three terms of \eq{dG} involve
photons linking different quantities; for example, the last term has one end of
a photon attached to $K$ and the other to $G$ in \eq{G}.

By formally solving \eq{G^mu_eq} we obtain an explicit expression for 
$G^\mu$ \cite{g4d}:
\be
\GG^\mu = \GG \Gamma^\mu\GG;\hspace{1cm}
\Gamma^\mu = \Gamma_0^\mu + \K^\mu,    \eqn{G^mu}
\ee
where
\be
\Gamma_0^\mu = \GG_0^{-1} \GG_0^\mu \GG_0^{-1}.
\ee
In the case of two strongly interacting particles, $G_0=d_1d_2$ where $d_i$ is
the dressed propagator for particle $i$ (with dressing due to strong
interactions only) and therefore
\be
G_0^\mu = d_1^\mu d_2 + d_1 d_2^\mu
\ee
where $d_i^\mu$ is given in terms of the one-particle electromagnetic vertex
function $\ggamma_i^\mu$ by
\be
     d_i^\mu = d_i \ggamma_i^\mu d_i,
\ee
and
\be
\Gamma_0^\mu =  \ggamma_1^\mu d_2^{-1} + d_1^{-1}\ggamma_2^\mu
\ee
is the electromagnetic current of the two-particle system in
impulse approximation.  $\K^\mu$ in \eq{G^mu} corresponds to the interaction current and can
also be obtained by the gauging of equations method if $K$ is given
nonperturbatively by an integral equation. In any case, here we shall treat
$K^\mu$ as an input quantity that has previously been constructed.

\eq{dG} is an integral equation for the electromagnetic corrections $\dG$ which
can be formally solved in just the same way that \eq{G^mu_eq} was solved to
get \eq{G^mu}. We obtain that
\be
\dG = \GG\Delta\GG
\ee
where
\be
\Delta= \dK + \GG_0^{-1}\dG_0\GG_0^{-1}
+\left(\Gamma^\mu\GG\Gamma^\nu-\Gamma_0^\mu \GG_0\Gamma_0^\nu\right) D_{\mu\nu}.
\eqn{Delta}
\ee
This is the central result of our paper. It expresses the complete set of lowest
order electromagnetic corrections to $G$ in terms of the solution to the initial
strong interaction problem ($G$ itself).  As discussed below, $\Delta$ forms the
complete set of electromagnetic corrections to the potential $K$. Because of
this completeness, calculations of physical quantities using \eq{Delta} will be
gauge invariant.\begin{figure}[t]
\hspace*{3mm}  \epsfxsize=15.5cm\epsfbox{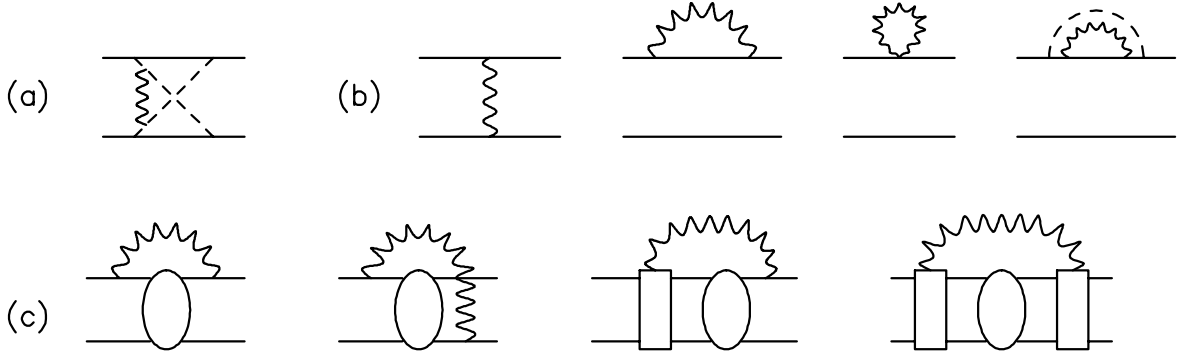}
\vspace{4mm}

\caption{\fign{ope} Diagrams contributing to the electromagnetic correction
$\Delta$ of \eq{Delta}. (a) An example of a correction belonging to $\delta K$
for the case where solid lines are nucleons and dashed lines are pions. (b) The
diagrams contributing to $G_0^{-1}\delta G_0 G_0^{-1}$: one-photon exchange, two
types of particle dressing by photon, and an example of a correction belonging
to $\delta\Sigma$.  (c) Contributions to $\left(\Gamma^\mu G
\Gamma^\nu-\Gamma_0^\mu G_0 \Gamma_0^\nu\right)D_{\mu\nu}$: the oval represents
the strong interaction Green function $G$ (with $G_0$ subtracted in the first
two diagrams), while the rectangle with attached photon represents $K^\mu$.  All
photon-particle vertices are dressed (by strong interactions).}
\end{figure}

The various contributions to $\Delta$ are illustrated diagramatically in
\fig{ope} for the two-body case (note however that \eq{Delta} is valid for any
number of quarks or hadrons).  To reveal the contributions to
$G_0^{-1}\dG_0G_0^{-1}$ we write
\be
\dG_0 = \delta(d_1d_2) = \delta d_1 d_2 + d_1\delta d_2 + G_0 K^\gamma G_0
\ee
where $\delta d_i$ is the complete set of electromagnetic corrections to $d_i$
and $K^\gamma$ is the one photon exchange potential as illustrated in the first
diagram of \fig{ope}(b). As the dressed quark or hadron propagator $d$ is
expressed in terms of the integral equation
\be
d = d_0 + d_0\Sigma d         \eqn{d}
\ee
where $d_0$ is the bare propagator and $\Sigma$ is the (strong interaction)
dressing term, we can determine $\delta d$ in just the same way that $\dG$ was
determined from \eq{G}. We obtain that
\be
d^{-1}\delta d\, d^{-1}\equiv\Sigma^\gamma
= \delta\ggamma_0+\delta\Sigma+\ggamma^\mu d\ggamma^\nu D_{\mu\nu}
\eqn{X}
\ee
where $\delta\ggamma_0$ is defined by the diagram in \fig{sigma}(a) and
corresponds to the term $\ggamma_0^{\mu\nu}D_{\mu\nu}$ where
$\ggamma_0^{\mu\nu}$ arises from the gauging of $\ggamma_0^\mu$. Note that
$\delta\ggamma_0=0$ for spin 1/2 particles but is not zero, for example, for
spin 0 particles. 
Diagrams contributing to the last two terms of \eq{X} are also illustrated in
\fig{sigma}.
\begin{figure}[b]
\hspace*{.5cm}  \epsfxsize=15.5cm\epsfbox{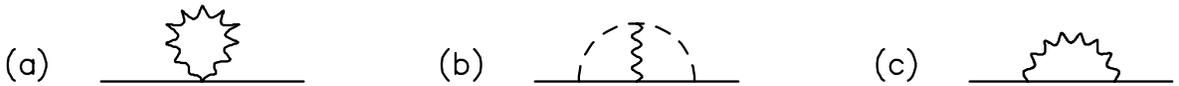}
\vspace{4mm}

\caption{\fign{sigma} Diagrams contributing to the complete set
of electromagnetic corrections of a single quark or hadron - see
\eq{X}:
(a) $\delta\ggamma_0$, (b) an example of a contribution to $\delta\Sigma$
in the case where the solid line is a nucleon and the dashed line is a pion, 
and (c) $\ggamma^\mu d \ggamma^\nu D_{\mu\nu}$.}
\end{figure}
Thus
\be
G_0^{-1}\dG_0G_0^{-1} = K^\gamma + \Sigma_1^{\gamma}d_2^{-1}
+ \Sigma_2^{\gamma}d_1^{-1}
\ee
as illustrated in \fig{ope}(b).

The last term of \eq{Delta}, $\Gamma_0^\mu \GG_0\Gamma_0^\nu D_{\mu\nu}$, is
made up of the first two diagrams of \fig{ope}(b). Its role in \eq{Delta} is
clearly to subtract those contributions from $\Gamma^\mu \GG\Gamma^\nu
D_{\mu\nu}$ which already appear in $\GG_0^{-1}\dG_0\GG_0^{-1}$. It follows that
$\left(\Gamma^\mu \GG\Gamma^\nu -\Gamma_0^\mu
\GG_0\Gamma_0^\nu\right)D_{\mu\nu}$ is just the connected part of $\Gamma^\mu
\GG\Gamma^\nu D_{\mu\nu}$ minus the one-photon exchange contributions -
as illustrated in \fig{ope}(c).

As noted previously, $\delta K$ does not have photons attached to external legs.
On the other hand, if we define the corrected Green function as
\be
G'\equiv G+\dG = G+G\Delta G,
\ee
we see that to order $e^2$ in the electromagnetic interaction
\be
G' \approx \left(G^{-1}-\Delta\right)^{-1}
=\left(G_0^{-1}-\K-\Delta\right)^{-1}
\ee
indicating that $\Delta$ can be considered as the full electromagnetic
correction to the potential $\K$. Defining $K'=K+\Delta_c$ where $\Delta_c$
is the connected part of $\Delta$, it is easy to see that
\be
G'=G_0'+G_0'K'G'    \eqn{G_full}
\ee
where $G'_0=d_1'd_2'$ with $d'=d+d\Sigma^\gamma d'$ being the fully dressed 
one-particle propagator with all second order
electromagnetic interactions included, and $K'=K + \delta K+ K^\gamma +
\left(\Gamma^\mu\GG\Gamma^\nu-\Gamma_0^\mu \GG_0\Gamma_0^\nu\right)D_{\mu\nu}$
is the corrected potential incorporating the complete set of (connected) second
order electromagnetic corrections, including those where photons are attached to
external legs. The gauge invariance of on-mass-shell amplitudes corresponding to
$G'$ is therefore guaranteed (an explicit proof will be given later in a more
detailed article). 

\section{Electromagnetic corrections to the bound state mass}

The strong interaction bound state wave function $\psi$ satisfies the bound
state equation
\be
\psi_P=G_0K\psi_P          \eqn{psi}
\ee
where the total momentum $P$ is on mass shell: $P^2=M^2$. The bound state wave
function also satisfies the normalization condition
\be
i \bar{\psi}_P
\frac{\partial}{\partial P^2}\left(G_0^{-1}-K\right)\psi_P = 1
\eqn{norm}
\ee
where the derivative is evaluated at $P^2=M^2$. The mass $M$ of this bound state
involves strong interactions only and will change by some amount $\delta M$ on
inclusion of the electromagnetic interaction. Our goal is to find $\delta M$ for
the case where all possible electromagnetic interactions are included up to
order $e^2$.

With the full Green function
including electromagnetic interactions given by \eq{G_full}, the bound state
equation becomes
\be
\psi_{P'}'=G_0'K'\psi_{P'}' \hspace{.5cm}\Rightarrow\hspace{.5cm}
\left.\left({G_0}\!^{-1}-K-\Delta\right)\right|_{P'^2=M'^2} \psi_{P'}'=0
\eqn{psi_full}
\ee
where $\psi'_{P'}=\psi_P+\delta\psi_P$ is the wave function with electromagnetic
corrections included, and $P'^2=M'^2$ where $M'=M+\delta M$ is the corresponding
bound state mass.  Making an expansion up to order $e^2$ around the point
$P^2=M^2$ and 
using both \eq{psi} and \eq{norm} we obtain that
\be
\delta M = \frac{1}{2M}\bpsi_P\Delta\psi_P     \eqn{delta_M}
\ee
with the right hand side evaluated at $P^2=M^2$. The gauge invariance of the
mass correction $\delta M$ as given by \eq{delta_M} follows from the
completeness of the included electromagnetic interactions, although again we
shall leave an explicit proof to a more detailed publication.
\bigskip

The authors would like to thank H.\ Leutwyler and M.P.\ Locher for their very
informative comments and the Australian Research Council for their financial
support.


\end{document}